\documentclass[twocolumn,prd,nofootinbib,aps,floats,floatfix,amsmath,amssymb,secnumarabic,preprintnumbers]{revtex4} %
\usepackage[final]{graphicx}
\usepackage[dvipsnames]{xcolor}
\usepackage{amsmath}
\usepackage{bbm}
\usepackage{amsfonts}
\usepackage{amssymb}
\usepackage{latexsym}
\usepackage{graphicx}
\usepackage[english]{babel}
\usepackage{multirow}
\usepackage{float}
\usepackage{url}
\usepackage{hyperref}
\usepackage{slashed}
\usepackage{xcolor} 
\usepackage[utf8]{inputenc}

\def\sfrac#1#2{{\textstyle{#1\over #2}}}
\newcommand{\be}{\begin{equation}}
\newcommand{\ee}{\end{equation}}
\newcommand{\ba}{\begin{array}}
\newcommand{\ea}{\end{array}}
\newcommand{\bea}{\begin{eqnarray}}
\newcommand{\eea}{\end{eqnarray}}
\newcommand{\sss}{\scriptscriptstyle}

\newcommand{\tr}{{\rm tr}}

\newcommand{\nn}{\nonumber}
\newcommand{\R}{{\sss R}}

\renewcommand{\H}{{\sss H}}

\renewcommand{\L}{{\sss L}}
\newcommand{\W}{{\sss W}}
\newcommand{\Z}{{\sss Z}}
\newcommand{\nada}{{}}

\begin{document} 
\preprint{CERN-TH-2017-136}
\title{$B$ decay anomalies from nonabelian local horizontal symmetry}

\author{James M.\ Cline}
\affiliation{CERN Theoretical Physics Department and McGill University, Department of Physics}
\author{Jorge Martin Camalich}
\affiliation{CERN Theoretical Physics Department, Geneva, Switzerland}

\begin{abstract}  
Recent anomalies in $B\to K^{(*)}\ell\ell$ meson decays are consistent with exchange of
a heavy $Z'$ vector boson.  Here we try to connect such new physics
to understanding the origin of flavor, by gauging generation number.  
Phenomenological and theoretical considerations suggest that the
smallest viable flavor symmetry (not including any extra U(1) factors)
is chiral SU(3)$_L\times$SU(3)$_R$, which acts only on generation indices and
does not distinguish between quarks and leptons.  Spontaneous breaking of the symmetry 
 gives rise
to the standard model Yukawa matrices, and masses for
the 16 $Z'$-like gauge bosons, one of which is presumed to be light
enough to explain the $B\to K^{(*)}\ell\ell$ anomalies.
We perform a  bottom-up study of this framework, showing
that it is highly constrained by LHC dilepton searches, 
meson mixing, $Z$ decays and CKM unitarity.
Similar anomalies are predicted for semileptonic
decays of  $B$ to lighter mesons, with excesses in the $ee,\tau\tau$
channels and deficits in $\mu\mu$, but no deviation in $\nu\nu$. 
The lightest $Z'$ mass is $\lesssim 6$\,TeV if the 
gauge coupling is $\lesssim 1$.
\end{abstract}
\maketitle

\section{Introduction}

Particle physicists have long been waiting for some definitive sign of
a breakdown in the Standard Model (SM), which generally works so well
as to recall Lord Kelvin's famous statement, ``There is nothing new to
be discovered in physics now. All that remains is more and more
precise measurement."   But it has also been anticipated that 
precision measurements, in the context of flavor, could be the most
likely harbinger of new physics (NP), since flavor changing neutral
currents (FCNCs) are so highly suppressed in the SM~\cite{Glashow:1976nt,Glashow:1970gm}.
The natural progression of such a signal would be a gradual accumulation of
tension in some flavor observables.  In recent years, tensions have
been mounting in {semileptonic $B$ decays}, which have been measured with
increasing accuracy {at the LHC}~\cite{Aaij:2014ora,Aaij:2015esa,Aaij:2015oid,Aaij:2015yra,Aaij:2017vbb,ATLAS:2017dlm}
and $B$ factories~\cite{Lees:2012xj,Lees:2013uzd,Huschle:2015rga,Abdesselam:2016xqt,Wehle:2016yoi} 
{and point to new sources of lepton-universality violation (LUV) in nature.}

In particular {for the $b\to s\ell\ell$ FCNC transitions}, LHCb has found compelling discrepancies in the ratios
\be
R_X = \frac{\mathcal{B}(\bar{B} \rightarrow X \, \mu^+ \,\mu^- )}
{\mathcal{B}(\bar{B} \rightarrow X \, e^+ \, e^-)}
\ee
for decays into $X=K,\,K^*$, 
which are predicted to be very close to 1 in the SM~\cite{Hiller:2003js,Bordone:2016gaq,Geng:2017svp}.
The measured values are $R_K = 0.745\, \pm0.09\,\pm 0.036$ \cite{Aaij:2014ora}, 
$2.6\sigma$ below the SM prediction, and 
$R_{K^*} = 0.660^{+0.110}_{-0.070}\,\pm 0.024$ (low $q^2$),
$R_{K^*} = 0.685^{+0.113}_{-0.069}\,\pm 0.047$, where $q^2$ is the 
invariant mass of the lepton pair \cite{Aaij:2017vbb}.  The significance
of the discrepancy in each bin is $2.2$-$2.5\,\sigma$.  Moreover an
angular analysis of $B\to K^*\mu\mu$  \cite{Aaij:2015oid} {suggests}
a $3.4\,\sigma$ discrepancy.

The quantity $R_X$ is particularly interesting because hadronic
uncertainties in the decay rate cancel to a high degree in the ratio,
making this a ``clean'' observable (see e.g.~\cite{Geng:2017svp}). Other measurements such as 
branching fractions and the $B\to K^*\mu\mu$ angular observables
mentioned above are not so 
theoretically clean, but it is interesting that their inclusion tends to
reinforce the evidence from clean observables
only~\cite{Geng:2017svp,DAmico:2017mtc,Altmannshofer:2017yso,Capdevila:2017bsm,Ciuchini:2017mik,Bardhan:2017xcc,
Neshatpour:2017qvi,Alok:2017jaf,Alok:2017su}, 
a further indication that the effect could be real.
The best fits are provided by NP contributions involving the effective operators 
\begin{align}
\label{eq:HweakSM}
\mathcal H_{\rm w}&\supset -\frac{\alpha_{\rm em}}{4\pi v^2}
\lambda_{bs}^{(t)}\left[C_{b_L\ell_L}(\mu)
\left(\bar s_\L \gamma_\mu b_\L\right) 
\left(\bar\ell_\L\gamma^\mu\ell_\L\right)\nonumber\right. \\
&\left.+C_{b_L\ell_R}(\mu)\left(\bar s_\L \gamma_\mu b_\L\right) 
\left(\bar\ell_\R\gamma^\mu\ell_\R\right)\right],
\end{align}
where $\lambda_{bs}^{(t)}=V_{tb}V_{ts}^*$ and the SM contributions
are $C_{b_L\ell_L}(m_b)=8.64$, $C_{b_L\ell_R}(m_b)= -0.18$~\cite{Bobeth:2003at}.
Since $|C_{b_L\ell_L}|\gg|C_{b_L\ell_R}|$, it is possible to fit the
data well with NP contributions  to the left-handed leptonic
operators $C_{b_L\ell_L}$ alone.

The $B\to K^{(*)}\ell\ell$ anomalies have inspired many model-building
efforts, with the most popular proposals involving exchange of 
heavy $Z'$ vector bosons~\cite{Gauld:2013qba,Gauld:2013qja,Buras:2013dea,Altmannshofer:2014cfa,
Crivellin:2015lwa,Sierra:2015fma,Crivellin:2015era,Celis:2015ara,Carmona:2015ena,Fuyuto:2015gmk,Chiang:2016qov,
Kim:2016bdu,Cheung:2016exp,Crivellin:2016ejn,GarciaGarcia:2016nvr,Datta:2017pfz,Ko:2017lzd,Alonso:2017bff,
Bonilla:2017lsq,Alonso:2017uky,Ellis:2017nrp,Tang:2017gkz,Chiang:2017hlj,Chivukula:2017qsi,King:2017anf}, 
leptoquarks~\cite{Hiller:2014yaa,Gripaios:2014tna,Allanach:2015ria,Alonso:2015sja,Bauer:2015knc,
Sahoo:2015pzk,Kumar:2016omp,Das:2016vkr,Li:2016vvp,Chen:2016dip,Becirevic:2016yqi,Becirevic:2016oho,
Mileo:2016zeo,Hiller:2016kry,Sahoo:2016pet,Popov:2016fzr,Barbieri:2016las,Chen:2017hir,Crivellin:2017zlb,Hiller:2017bzc,
Cai:2017wry,Becirevic:2017jtw,Chauhan:2017ndd} 
or loop-induced transitions~
\cite{Gripaios:2015gra,Bauer:2015knc,Arnan:2016cpy,Kamenik:2017tnu,Das:2017kfo,He:2017osj}. The data can be well
fit in simplified models that are designed to address only
$R_{K^{(*)}}$, but one naturally hopes that the complete picture would
shed greater light on one of the biggest puzzles of the SM, the origin
of flavor.  If flavor symmetry is local and spontaneously broken, then
heavy $Z'$ gauge bosons would inevitably arise, possibly having
couplings with
the right flavor structure for explaining the anomalies~\cite{Altmannshofer:2014cfa,
Crivellin:2015lwa,Alonso:2017bff,Alonso:2017uky}.
This is the approach we take, with the goal of adopting the 
smallest nonabelian flavor
symmetry group that seems to be consistent with the observations,
while fully accounting for the structure of the SM Yukawa matrices.

The simplest possibility for a generational symmetry as the origin of
flavor  would be to couple all SM fermions vectorially to a single
SU(3)$_\H$ generation group. Although global  fits to $B\to
K^{(*)}\ell\ell$ decays disfavor purely vectorial currents to the
quarks, it was noted in ref.\ \cite{Alonso:2017bff} that chiral
currents can arise for the flavor-changing transitions if only
left-handed quarks need to be rotated when diagonalizing the quark
masses; the full flavor symmetry group must include a U(1)$_{B-L}$
factor to account for neutrino masses in this model.

Here we consider a different possibility, by assuming the larger
chiral group SU(3)$_L\times$SU(3)$_R$ with no U(1) factor.  
In addition, we attempt to give a detailed account of the
origin of the SM fermion masses  within the same framework,
as explained in section \ref{sec:model}. It turns
out to be  highly constrained, with phenomenological requirements
restricting the model-building choices at almost every step.  We make  a number of predictions for collider searches and
precision studies that are imminently testable, as explained in section
\ref{sec:constraints}.  Further consequences of the model, 
focusing on
physics above the scale needed to explain $R_{K^{(*)}}$, are discussed
in section \ref{sec:UV}.  We summarize the distinctive features of
our model and its differences with previous proposals in section
\ref{sec:conc}.
Appendix \ref{app:LV} presents the constraints on possible lepton
flavor violation that may be present in the model, while appendix
\ref{app:vector} explains why a simpler related model, with vectorial
SU(3)$_\H$ flavor group and no U(1) factor, is not viable.

\section{Model}
\label{sec:model}

In order to generate the SM fermion masses and to cancel anomalies,
we add a set of fermions $U_{\L,\R}$, $D_{\L,\R}$, $E_{\L,\R}$, $N_{\L,\R}$
and scalar fields $\Phi_{u,d,l,\nu}$, ${\cal M}$, $\Phi_6$, that transform
as shown in in table \ref{charges},
\bea
\label{eq:model}
{\cal L}_{\rm yuk} &=& 	
{\lambda_u}\bar Q_\L \tilde{H}U_\R + \lambda'_u \bar U_\L\Phi_u
u_\R  + \lambda''_u\,\bar U_\L {\cal M} U_\R\nn\\
 &+& 
{\lambda_d}\bar Q_\L {H}D_\R + \lambda'_d\bar D_\L\Phi_d d_\R
	+ \lambda''_d\,\bar D_\L {\cal M} D_\R\nn\\
	&+& {\lambda_l}\bar L_\L {H}E_\R 
	\ + \lambda'_l\bar E_\L\Phi_l l_\R \ \,+ 
\lambda''_l\,\bar E_\L {\cal M} E_\R\nn\\
	&+& {\lambda_\nu}\bar L_\L \tilde{H}N_\R +
	\lambda'_\nu\bar N_\L \Phi_\nu \nu_\R +
\lambda''_\nu\, \bar N_\L {\cal M}
N_\R\nn\\
	&+& \lambda_6\, \nu_\R^T\Phi_6\,\nu_\R 
\eea
The new fermions play a double role, by cancelling the anomalies
of SU(3)$_L\times$SU(3)$_R$, and by generating the SM Yukawa couplings. 
A $Z_2$ symmetry under which the right-handed SM fermions and $\Phi_f$ are charged 
prevents direct flavor-universal mass terms such as $\bar U_\R u_\R$.
The scalars $\Phi_{u,d,l,\nu}$, ${\cal M}$, $\Phi_6$ 
are present 
to spontaneously break this symmetry and to dynamically generate  the flavor 
structure of the SM, as we now show.  

For simplicity, we take ${\cal M}$ to get VEVs proportional to the
unit matrix
\be
\label{eq:Mdiags}
	\lambda''_f\langle {\cal M}\rangle = M_f\cdot{\mathbf 1}
\ee
 while $\langle\Phi_{f}\rangle$ may be more
complicated.  We further assume that ${M}_{f}\gg
\lambda'_f\langle\Phi_{f}\rangle$, with the possible exception
of $f=u$ because of the large top quark mass.  On the other hand
$\langle\Phi_6\rangle$ is much greater than
the other VEVs, so that the right-handed neutrinos are very heavy.

Integrating out the heavy fields gives rise to the dimension-5 
and 7 operators
\bea
{\cal L}_{\rm yuk} &=& 	
{1\over \Lambda_u}\bar Q_\L \tilde{H}\Phi_u u_R + 
{1\over \Lambda_d}\bar Q_\L {H}\Phi_d d_R
+{1\over \Lambda_l}\bar L_\L {H}\Phi_l l_R\nn\\
	&+& 
 {1\over \Lambda_\nu^2}(\bar L\tilde H)\Phi_\nu 
{1\over \lambda_6\langle\Phi_6\rangle}\Phi_\nu^T(\tilde H\bar L)^T
	+{\rm h.c.}
\label{yuk}
\eea
that become the fermion mass matrices.  The mass scales in the
denominators are given by $\Lambda_f^{-1} = 
\lambda_f\lambda_f'/{M}_f$, except possibly for $f=u$ where we use the
more exact expression
\begin{align}
\label{eq:exactmassexp}
\Lambda_u^{-1} = \frac{\lambda_u\lambda_u'}{\sqrt{{
M}_u^2 + (\lambda_u'\langle\Phi_u\rangle)^2}}. 
\end{align}
Below, we will assume
that $\langle\Phi_u\rangle$ is a diagonal matrix, and only the 
$\langle\Phi_u\rangle_{33}$
element will be large enough to matter in this more exact expression.

\begin{table}[b] 
\begin{tabular}{|c|c|c|c|c|c|c|}
\hline
Field  & U(1)$_y$ & SU(2)$_\L$ & SU(3)$_c$ & SU(3)$_\L$ & SU(3)$_\R$ & $Z_2$\\
\hline
$Q_\L$   & $\phantom{+}\sfrac16$  & $2$   & $3$   &  $3$   & $1$ & $\phantom{+}1$  \\
$L_\L$   &  $-\sfrac12$   &  $2$  &  $1$ &  $3$     & $1$ & $\phantom{+}1$ \\
$u_\R$ & $\phantom{+}\sfrac23$  &  $1$  & $3$ &  $1$   & $3$ & $-1$ \\
$d_\R$ & $-\sfrac13$  &  $1$      &  $3$  & $1$        & $3$ & $-1$   \\
$e_\R$ & $-1$  &  $1$   &  $1$   &  $1$         & $3$ & $-1$  \\
$\nu_\R$ &  $\phantom{+}0$ &  $1$  &  $1$    &  $1$  & $3$  & $-1$          \\
\hline
$\Phi_u,\Phi_d,\Phi_e,\Phi_\nu$ &   $\phantom{+}0$  &  $1$    &  $1$ &
  $1$ & $8$ & $-1$\\
${\cal M}$ &   $\phantom{+}0$  &  $1$    &  $1$ &
  $3$ & $\bar 3$  & $\phantom{+}1$\\
$\Phi_{8,1}$,  $\Phi_{8,2}$       &  $\phantom{+}0$ & $1$  &  $1$   &
$8$ & $1$& $\phantom{+}1$\\  
$\Phi_6$ & $\phantom{+}0$ & $1$  & $1$ & $1$ & $6$ & $\phantom{+}1$\\
$U_\R$ & $\phantom{+}\sfrac23$  &  $1$ &  $3$     &  $3$     &    $1$  & $\phantom{+}1$ \\
$D_\R$ & $-\sfrac13$  &  $1$  &  $3$    &  $3$      &    $1$ & $\phantom{+}1$\\
$E_\R$ & $-1$  &  $1$      &  $1$   & $3$          & $1$ & $\phantom{+}1$\\
$N_\R$ & $\phantom{+}0$ &  $1$  &  $1$    &  $3$  &    $1$        & $\phantom{+}1$\\
$U_\L$ & $\phantom{+}\sfrac23$  &  $1$ &  $3$     &  $1$     &    $3$   & $\phantom{+}1$\\
$D_\L$ & $-\sfrac13$  &  $1$  &  $3$    &  $1$      &    $3$ & $\phantom{+}1$\\
$E_\L$ & $-1$  &  $1$      &  $1$   & $1$          & $3$ & $\phantom{+}1$\\
$N_\L$ & $\phantom{+}0$ &  $1$  &  $1$    &  $1$  &    $3$       & $\phantom{+}1$ \\
\hline
\end{tabular}
\caption{Field content and charges of model.  The first three lines
are the
SM fermions, including right-handed neutrinos, while the following
contain the new field content.}
\label{charges}
\end{table}

\subsection{Fermion and gauge boson masses}
\label{fgbm}

After the $\Phi_f$ scalars and the Higgs field get VEVs, the
SM fermions get Dirac masses
\be
	(m_f)_{ij} = { v\over \Lambda_f} 
\langle\Phi_f\rangle_{ij}
\label{fmass}
\ee
(with $\Lambda_u$ taken to be a matrix as explained above)
while the neutrinos get the Majorana mass matrix
\be
	\tilde m_\nu = {v^2\over\Lambda_\nu^2}\,
	\langle\Phi_\nu\rangle\,
{1\over \lambda_6 \langle\Phi_6\rangle}\,\langle\Phi_\nu\rangle^T
\ee
with $v\langle\Phi_\nu\rangle/\Lambda_\nu$ playing the role of the 
Dirac mass matrix in the seesaw formula.

The large VEV $\langle\Phi_6\rangle$ breaks SU(3)$_\R\times$SU(3)$_\L
\to$SU(3)$_\L$ at a high scale, so we henceforth ignore the heavy gauge
bosons associated with SU(3)$_\R$.\footnote{ In general the mass eigenstates are mixtures of 
$A_\L$ and $A_\R$, but if $\langle\tilde\Phi_6\rangle \gg 
\langle{{\cal M}}\rangle$ as assumed, then the lightest 8 of the 16 gauge bosons will be
mostly $A_\L$, with a very small admixture of $A_\R$.  For simplicity
we will henceforth consider $A_\R$ to be decoupled and ignore this
small mixing.}\  
The terms that give masses to the SU(3)$_\L$ gauge bosons are
\bea
\label{gbmass}
{\cal L}_{gb} &=& 
g_\L^2\,\sum_{i=1,2}\tr\left([\Phi^\dagger_{8,i},A_\H][A_\H,\Phi_{8,i}]
\right)\nn\\
 &+& 
	g_\L^2\, \tr\left({\cal M}^\dagger{{\cal M}}
A_\L^2 \right)
\eea
where $A_{\L}^{{\mu}} = T^A A_{\L,A}^\mu$ with generators of the fundamental
representation, and $g_{\L}$ is the SU(3)$_{\L}$ gauge coupling. 
We will show that the $B$ decay anomalies motivate us
to further break SU(3)$_\L\to$ {U(1)$_{8}$}, the U(1) subgroup
whose gauge boson $Z' = A^{8}_\L$ couples to the diagonal 
generator $T^8$. 
This is the reason for including the $\Phi_{8,i}$ octet scalars.  
It suffices to have VEVs of the form $\langle\Phi_{8,1}\rangle =\alpha T^1$,
$\langle\Phi_{8,2}\rangle =\beta T^2$, with $\alpha,\beta\gg {\rm
TeV}$ to give large masses to all components of $A_\L$ except $A_{\H,8}$ 
as desired. 

The identification of $T^8$ as a special direction in the space of
generators implies a choice of basis for the fermion flavors.  We
are assuming that in this basis, the mass matrices of the quarks
and charged leptons are diagonal, in the limit where CKM mixing is
neglected.  To include CKM mixing, we will make the simplifying
assumption that the up-like mass matrix 
$(m_u)_{ij}$ is diagonal,
and all the mixing comes from $(m_d)_{ij}$.  This
choice is particularly convenient for revealing that our model enjoys
the properties of minimal flavor violation (MFV) 
\cite{Chivukula:1987py,DAmbrosio:2002vsn}; all the FCNCs that arise
from $Z'$ exchange
explicitly have the same CKM structure as in the SM.

We emphasize that the assumption of $(m_u)_{ij}$
being diagonal is not crucial to the more general framework presented
here.  It would also be consistent to have off-diagonal contributions
to $(m_u)_{ij}$ similar in relative size to those in $(m_d)_{ij}$.  For
example, suppose that the fermion masses are diagonalized as usual by unitary transformations
$f_\L \to V_f^{\L\dagger} f_\L$, $f_\R \to V_f^{\R\dagger} f_\R$,
such that $V_u^{\L} = V_d^{\L\dagger} = \sqrt{V_{\rm CKM}} \equiv
{\mathbf 1} + \sfrac12 \delta V - \sfrac18\delta V^2+\dots$, where
$\delta V = V_{\rm CKM} - {\mathbf 1}$.  Then the predictions we
present in the following would be similar to those in the simpler case
where $V_u^{\L} = {\mathbf 1}$, $V_d^{\L} =  V_{\rm CKM}^\dagger$.
The flavor-changing couplings of $Z'$ to down-type quarks would be
approximately half as large, this amount being shifted into those of the up-like quark sector.  Detailed
predictions would change but the overall picture, including MFV
structure, would be preserved.  We defer the study of
such generalizations for possible future work.

\subsection{Currents}

Diagonalizing the gauge boson mass matrix determines the mass
eigenstates as $\hat A_\L^a = O_{aB} A_\L^B$ where $O$ is an orthogonal
matrix.  Our model is such that $Z' \cong \hat A_\L^8$ is the 
lightest gauge boson, whose exchange is the origin of anomalous 
$B\to K^{(*)}\mu^+\mu^-$ decays. In general, a $Z'$ that has only
flavor-diagonal couplings could couple to the linear combination
of generators
\be
O_{1A} T^A \cong T^8 + {\epsilon\over\sqrt{3}} T^3 = 
\sfrac{1}{2\sqrt{3}} \left(\begin{array}{ccc} 1 + \epsilon & 0 & 0\\
	0 & 1 - \epsilon & 0 \\
	0 & 0 & -2 \end{array}\right)
\label{qgen}
\ee
(which must be traceless since they belong to SU(3)).  
It turns out
that, to avoid large FCNC's affecting $K$-$\bar K$ mixing, $\epsilon$
must be negligibly small.  Such operators, with complex coefficients,
are directly induced by exchange of the  $A^{1,2}_{\L}$
gauge bosons coupling to $T^{1,2}$, which constrains their masses to
be at the scale  $g_\L\langle\Phi_{8,i}\rangle\gtrsim
10^4\,$TeV.  Diagonalization of the gauge boson mass matrix then 
reveals that $\epsilon \sim {\cal M}^2/\Phi_8^2 \lesssim 10^{-8}$,
since $m_{Z'}\cong g_\L {\cal M}$ is at the TeV scale.
We therefore ignore $\epsilon$ in the following.

The fermion masses are diagonalized as usual by unitary transformations
$f_\L \to V_f^{\L\dagger} f_\L$, $f_\R \to V_f^{\R\dagger} f_\R$.  Then couplings of
$Z'$ to fermions in the mass basis are given by
the left-handed currents, 
\be
	 g_{\L}Z'_\mu\,\bar f_{\L}\, [ V_f^{\L} T^8
V_{f}^{\L\dagger}]\gamma^\mu f_{\L},
\label{current}
\ee
where by our simplifying assumption $V_u^{\L}={\mathbf 1}$ and
the CKM mixing is entirely due to $V_d^\L = V_{\rm CKM}^\dagger$.

Then as discussed in section \ref{fgbm}, flavor mixing in the 
down-quark sector has a  
structure resembling the MFV hypothesis.
The diagonal couplings to (left-handed) quarks are given by 
\be
	{g_\L\over 2\sqrt{3}}\left\{\begin{array}{rl}
	1,& u,d,c,s\\
	-2, & b,t\end{array}\right.
\label{cdiag}
\ee
while the off-diagonal ones are
\be
	-{\sqrt{3}\over 2}g_\L\,\left\{\begin{array}{rl}
	V_{ts}V_{td}^*,& s\to d\\
	V_{tb}V_{td}^*,& b\to d\\
	V_{tb}V_{ts}^*, & b\to s\end{array}\right\}
\label{codiag}
\ee

For the left-handed leptons, we require 
that $V_l T^8 V_l^\dagger$ is
nearly diagonal, to avoid tree-level lepton flavor changing neutral
currents.  Moreover the diagonal elements must violate flavor
universality to explain the $R_{K^{(*)}}$ anomalies. 
We assume that 
\be
	V_l^\L\,T^8\, V_l^{\L\dagger} \cong
\sfrac{1}{2\sqrt{3}}\left(\begin{array}{ccc}1 & \phantom{-}\epsilon_1 &
\epsilon_2\\
	\epsilon_1^* & -2 & \epsilon_3 \\
	 \epsilon_2^* & \phantom{-}\epsilon_3^* & 1 \end{array}\right)
\label{ltrans}
\ee
which  preserves the eigenvalues of (\ref{qgen})
for $\epsilon_i \ll 1$. Hence $V_l$ is
approximately of the form
\be
	V_l^\L\cong
\left(\begin{array}{ccc}0 & 1 & 0\\
	0 & 0 & 1\\ 1 & 0 & 0\end{array}\right)
\label{Vleq}
\ee
which is just a permutation, and has determinant $+1$.  This is
the unique SU(3) transformation that takes the generator $T^8$ into a 
diagonal form in which muons couple more strongly than electrons,
as indicated by the $R_{K^{(*)}}$ anomaly (and having the  right sign, as will be
established below), hence (\ref{Vleq}) is forced upon us. The
right-handed rotation $V_l^\R$ is still unconstrained, while the
transformation $V_\nu^\L$ is now determined in terms of the PMNS
neutrino mixing matrix $U$, as $V_\nu^\L =  U V_l^\L$.
For simplicity we will impose lepton flavor conservation by taking
$\epsilon_i = 0$ in the remainder.  If one relaxes this assumption,
the experimental constraints on lepton flavor violation require
that $|\epsilon_1| < 0.007$, $|\epsilon_{2,3}| < 0.7$, as shown in 
appendix \ref{app:LV}.

It is worth emphasizing that, while there is considerable freedom in 
choosing VEVs of the $\Phi_f$ fields to obtain the flavor structure
of the quark and lepton $Z'$ currents, there is also an important
restriction: the generators are traceless, forcing $Z'$ to couple with
similar strength to all quarks and leptons, in addition
to the phenomenologically motivated {$b\to s\ell\ell$} coupling.  This
leads to interesting constraints and predictions as we now explore.

%---------------------------------------------------------------------------------------------------
%
\begin{figure}[t]
\hspace{-0.4cm}
\centerline{
\includegraphics[width=0.95\hsize]{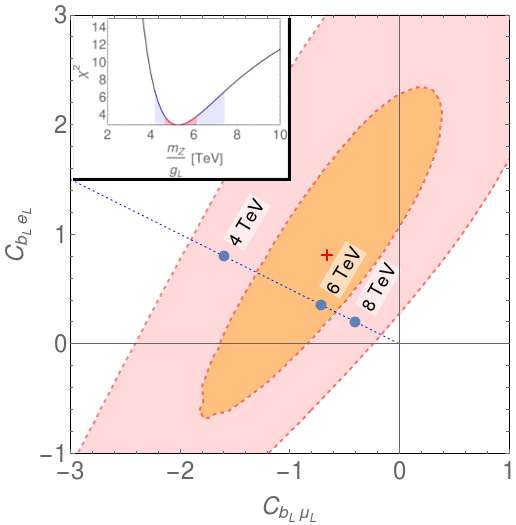}}
\caption{1$\sigma$ and 3$\sigma$ bounds given by the measurements of $R_{K^*}$, $R_K$ and the branching fraction of $B_s\to\mu\mu$ 
in the ($C_{b_L\mu_L}$, $C_{b_Le_L}$) plane. The line corresponds to the trajectory of the contributions of the $Z^\prime$ 
to these Wilson coefficients in our model, 
where we have indicated benchmark points of $m_{Z'}/g_L$, whereas the embedded plot shows the projected $\chi^2$. 
The absolute minimum in the plane, indicated by a red cross corresponds to $\chi^2_{\rm min}=2.64$. }
\label{RKsbound}
\end{figure}
%
%---------------------------------------------------------------------------------------------------

\section{Constraints and predictions}
\label{sec:constraints}

Having defined the model, there is only one combination of
parameters, $g_\L/m_{Z'}$, that is left to fit the 
$R_{K^{(*)}}$ anomalies.  Once this is done, a number of predictions
for related FCNC semileptonic meson decays, 
{neutral} meson oscillations, $Z$-decays, and violation of unitarity
of the CKM matrix follow. In addition a level of dilepton pair
production at the LHC is predicted that is close to current
constraints.  We discuss these issues in the following.

\subsection{Explaining the $R_{K^{(*)}}$ anomalies}

The contributions to $b\to s\ell\ell$ processes from purely semileptonic operators in the SM are contained in
eq.\ (\ref{eq:HweakSM}),
where the Wilson coefficients are 
independent of the lepton flavor.
Global fits to $R_{K^{(*)}}$ point to new lepton flavor 
nonuniversal contributions to these operators and including other $b\to s\mu\mu$ data suggests that part of this NP 
appears in the muonic operators. Contributions to other operators, 
such as those involving $b_R$ or different Lorentz structures, are disfavored as discussed in refs.~\cite{Geng:2017svp,DAmico:2017mtc,Alonso:2014csa}.     

From (\ref{codiag}) and (\ref{ltrans}) it follows that our model produces lepton-specific contributions precisely to $\mathcal O_{b_L\ell_L}$,
\begin{align}
\label{eq:predRKs}
	\delta C_{b_L\mu_L} =-2 \delta C_{b_Le_L}= -2 \delta C_{b_L\tau_L}= -{g_\L^2\over m_{Z'}^2}{2\pi v^2\over \alpha}
\end{align}
The CKM coefficient $\lambda_{bs}^{(t)}$ of this contribution has factored out with the SM normalization in 
eq.~(\ref{eq:HweakSM}), which is a consequence of the MFV-like structure of the $Z^\prime$ couplings to the quarks. 

In fig.~\ref{RKsbound} we show the trajectory of our model as a function of $m_{Z'}/g_L$ in the ($C_{b_L\mu_L}$, $C_{b_Le_L}$) plane 
compared to the best fit point to $R_K$, $R_{K^*}$ and $B_s\to\mu\mu$ of ref.~\cite{Geng:2017svp}. In the lower-left corner we also show the projection 
of the $\chi^2$ along $m_{Z'}/g_L$. As one can see, our model gives an excellent fit to the data, with a $\chi^2=2.8$ for 3 degrees of freedom, 
which represents a 4.22\,$\sigma$ improvement over the SM. The best fit point and 1\,$\sigma$ error interval is
\begin{align}
\label{eq:bestfit}
{m_{Z'}\over g_\L}=5.3^{+0.9}_{-0.6}~\text{TeV}.
\end{align}
{We have checked that adding the angular observables of $B\to
K^*\mu\mu$ in a global fit slightly narrows the constraint on
$C_{b_L\mu_L}$ but does not have a significant impact on the best
solution or improvement with respect to the SM.}\footnote{In
ref.~\cite{DAmico:2017mtc} at fit was performed for models that are
similar to ours, which gives a significantly
stronger bound on $C_{b_L eL}$.  This stems from their inclusion of
two data points whose respective preferred solutions 
for the minimum of $\chi^2$ are regions of parameter
space with small overlap; these are inclusive $B\to X_s e e$ and $B^+\to K^+ ee$. 
We do not include these observables in our fit (\ref{eq:bestfit});
doing so we find that the improvement in $\chi^2$ is comparable
and the best fit value is shifted to $m_{Z^\prime}/g_\L\simeq5.1$\,
TeV. We thank Guido D'Amico and Marco Nardecchia for discussions
clarifying this point.}

Eq.\ (\ref{eq:predRKs}) predicts 
excesses for the branching ratios of 
 $B\to K^{(*)}\tau\tau$,  $B\to K^{(*)}ee$, which are approximately half the deficit
in $B\to K^{(*)}\mu\mu$. A further consequence of the MFV couplings to the quarks is that similar effects should be measured in $B\to Ml^+l^-$, 
where $M$ is a zero-strangeness meson. In particular we predict 
\begin{align}
\label{eq:predRpi}
R_{\pi}\simeq R_{K},~~~~~ R_{\rho}\simeq R_{K^*},
\end{align}
for the decay channels with pions and $\rho$ mesons.\footnote{
Large differences in form factors between the channels could in
principle modify this prediction, but such differences are 
disfavored by approximate
SU$(3)$-flavor symmetry in the light-quark sector of QCD, and 
by explicit calculations~\cite{Straub:2015ica}.}

\bigskip
\subsection{$d_i\to d_j\bar\nu\nu$ decays}

Along  with the charged leptons, our $Z^\prime$ couples to neutrinos and hence contributes to rare decays such as $B\to K^{(*)}\bar \nu\nu$ and $K\to\pi\bar\nu\nu$. 
Interestingly, the former will be searched by Belle II and the latter will be better measured by the NA62 experiment in the coming year. In the SM 
the $d_i\to d_j\bar\nu\nu$ decays are induced by the low-energy operator
\begin{align}
\label{eq:Hweakneut}
\mathcal H_{\rm w}\supset  -\frac{\alpha_{\rm em}}{4\pi v^2}\lambda_{ij}^{(t)} C_{\nu_{\ell}}(\bar d_j\gamma_\mu d_{iL})(\bar \nu_\ell\gamma_\mu\nu_{\ell L}),
\end{align}
where $C_{\nu_{\ell}}\simeq-12.7$~\cite{Brod:2010hi}. The contributions of the $Z^\prime$ are  
\begin{align}
\label{eq:contrareneut}
\delta C_{\nu_{e}} = \delta C_{\nu_{\mu}} =-\delta C_{\nu_{\tau}}/2={g_\L^2\over m_{Z'}^2}{\pi v^2\over \alpha}=0.37.
\end{align}
Even though large deviations are predicted for decays into individual
neutrino flavors, what the experiments observe are the ``invisible'' $B\to K^{(*)}$ and $K\to\pi$ rates, in which the absolute contributions from the neutrino flavors are
summed over. An important consequence of the tracelessness of the current, eq.~(\ref{cdiag}), together with the fact that the matrix element contributing to this process is 
the same for all neutrino flavors, is that the net interference of 
the $Z^\prime$ and SM contributions vanishes. The NP 
contribution to the branching fraction is
thus given by the quadratic terms,
\begin{align}
\frac{\text{BR}^{\rm Z^\prime}}{\text{BR}^{\rm SM}}=\frac{|\delta C_{\nu_e}|^2+|\delta C_{\nu_\mu}|^2+|\delta C_{\nu_\tau}|^2}{3 |C_{\nu_\ell}|^2}\simeq 2\times 10^{-3}, 
\end{align}
an effect that will be hardly detectable in forthcoming experiments.  

\bigskip
\subsection{$\Delta F=2$ transitions}

\begin{figure}[t]
\hspace{-0.4cm}
\centerline{
\includegraphics[width=0.95\hsize]{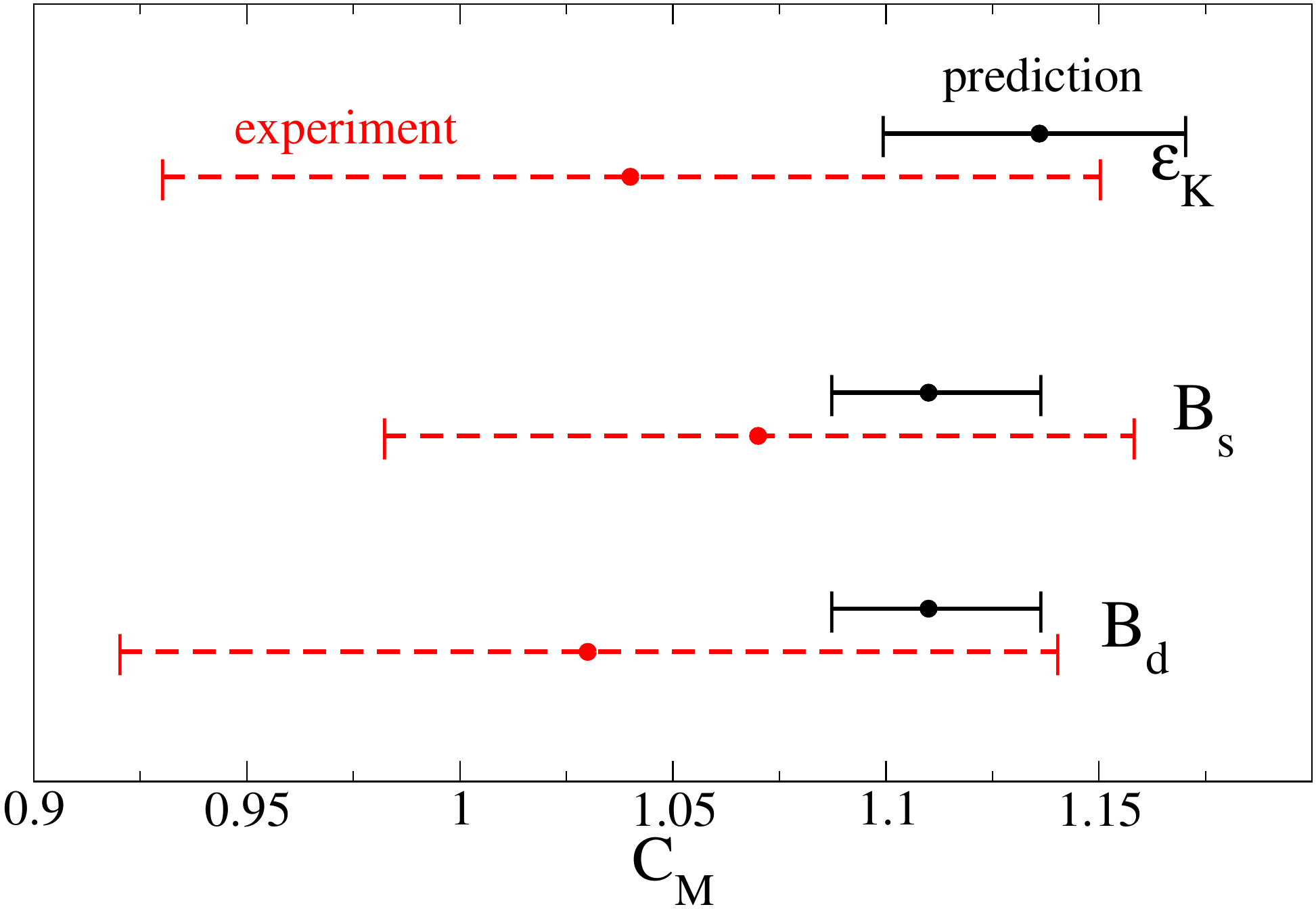}}
\caption{Predictions and experimentally allowed ranges for the
neutral meson mixing parameters (\ref{eq:CeK},\ref{eq:CBq}).}
\label{meson}
\end{figure}

Neutral-meson mixing receives tree-level contributions from $Z^\prime$-exchange in our model, yielding
\begin{align}
\label{eq:Hammixing}
\delta\mathcal H_{\rm w}=\frac{3g_L^2}{4m_{Z'}^2}
(\lambda_{ij}^{(t)})^2 (\bar d_{j}\gamma^\mu d_{\L i})
(\bar d_{j}\gamma_\mu d_{\L i}),
\end{align}
which has the same operator structure and combination of CKM matrix elements as the box diagram of the top quark in the SM. Parametrizing the deviation from the SM of 
the $\epsilon_K$ parameter in $K$-$\bar K$ mixing by~\cite{Bona:2007vi}
\begin{align}
\label{eq:CeK}
C_{\epsilon_K}=\frac{\text{Im}\langle  K^0|\mathcal H_{\rm w}|\bar K^0\rangle}{\text{Im}\langle  K^0|\mathcal H_{\rm w}^{\rm SM}|\bar K^0\rangle},
\end{align}
we obtain $C_{\epsilon_K}=1.14\pm0.04$ using eq.~(\ref{eq:bestfit}),
while the current experimental constraint is
$C_{\epsilon_K}=1.04\pm0.11$ at 1\,$\sigma$~\cite{Bona:2007vi}; 
the latter 
sets the lower bound $m_{Z'}/g_L\geq 5.1$ TeV, {which is quite close to} 
our best fit value (\ref{eq:bestfit}). In the case of 
$B_q$-$\bar B_q$ mixing the SM contribution is dominated by the 
top-loop diagram and its weak phase is aligned with that of 
the $Z^\prime$. Thus only the mass differences $\Delta m_{B_q}$
are {constraining}, which can be parametrized by
\begin{align}
\label{eq:CBq}
C_{B_q}=\left|\frac{\langle B_q|\mathcal H_{\rm w}|\bar B_q\rangle}{\langle B_q|\mathcal H_{\rm w}^{\rm SM}|\bar B_q\rangle}\right|. 
\end{align}
We obtain $C_{B_q}=1.12\pm0.03$ which is within the experimental limits $C_{B_s}=1.070\pm0.088$
and $C_{B_d}=1.03\pm0.11$~\cite{Bona:2007vi} and gives the
slightly weaker bound $m_{Z'}/g_L\geq4.8$ TeV.  The predictions
for $C_i$ and the experimental constraints are summarized in fig.\ 
\ref{meson}.  

%---------------------------------------------------------------------------------------------------
%
%
%---------------------------------------------------------------------------------------------------

\begin{figure}[t]
\hspace{-0.4cm}
\centerline{
\includegraphics[width=\hsize]{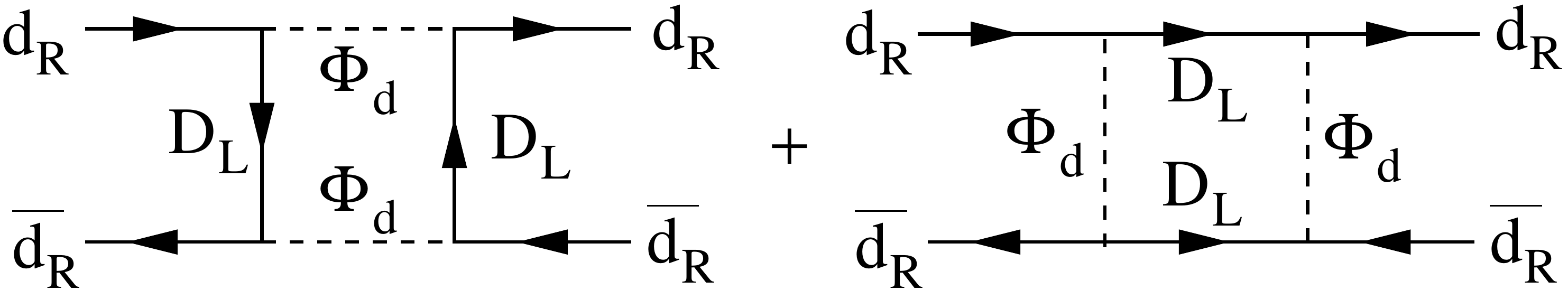}}
\caption{Loop contribution to neutral meson mixing.}
\label{box}
\end{figure}

There is a potentially dangerous contribution to $K$-$\bar K$ mixing
from the loop diagram \ref{box} from exchange of the heavy $\Phi_d$ 
and $D$ particles.  In the limit where all the states of the $\Phi_d$
octet are degenerate, {the contribution to the amplitude 
$(\bar d_{j}\gamma^\mu d_{\R,i})(\bar d_{j}\gamma^\mu d_{\R,i})$
is proportional to the product of SU(3) generators,
\begin{align}
 \sum_{r,s}\sum_{A,B}(T^A)_{ri}(T^A)_{js}(T^B)_{jr}(T^B)_{si}=0,~~~~~\text{for $i\neq j$}.
\end{align}
%  it gives rise to a flavor-diagonal operator
% $(\bar d_{\R,i}\gamma^\mu d_{\R,i})(\bar d_{\R,j}\gamma^\mu d_{\R,j})$
% that does not contribute to mixing.
} However if there are mass
splittings, then FCNCs get generated.  For example if $\Phi^1_d$
which couples to $T^1$ has mass-squared splitting $\delta m_\Phi^2$,
we find that the operator relevant to $K$-$\bar K$ mixing is
\be
	{{\lambda_d'}^4\, \delta m_\Phi^2\over 196\pi^2 m_\Phi^4}
	\,(\bar d_{\R}\gamma^\mu s_{\R})^2
\ee
Since the coefficient is real, it is constrained at the level
of $1/(10^3{\rm\,TeV})^2$ \cite{Bona:2007vi}.  We do not predict
the masses $m_\Phi$ or splittings $\delta m_\Phi^2$ here;
it would require constructing the full potential of the scalars which
is beyond the scope of this work.  Nothing ostensibly precludes choosing
$\delta m_\Phi^2/m_\Phi^4$ to be sufficiently small.

%---------------------------------------------------------------------------------------------------
%
%
%---------------------------------------------------------------------------------------------------

\subsection{Collider searches for resonant and nonresonant
dileptons}

\begin{figure}[t]
\hspace{-0.4cm}
\centerline{
\includegraphics[width=0.95\hsize]{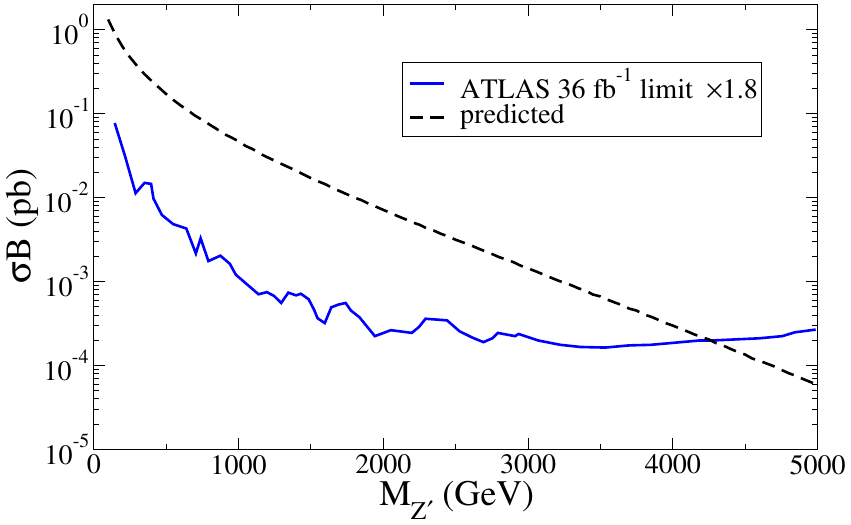}}
\caption{Production cross section for $Z'$ times branching ratio into
muons, versus $m_{Z'}$, as predicted by model and as constrained by
ATLAS \cite{ATLAS:2017wce}.  The rescaling of the limit by 1.8
corrects for the low branching ratio into $ee$ in our model; see text.}
\label{dilepton}
\end{figure}

A crucial test comes from the search for resonant production
of $Z'$ that decays to $\mu^+\mu^-$ and $e^+e^-$ \cite{ATLAS:2017wce}.  In our 
model, production occurs from all flavors of quarks in the proton
(but is dominated by the $u,d$ contributions),
according to the couplings (\ref{cdiag}).  The branching ratio for
decays into muons (electrons) is $B = \frac{1}{12}\, (\frac{1}{48})$, from (\ref{cdiag}) and
(\ref{ltrans}). Using MadGraph~\cite{Alwall:2014hca} to predict the resulting production
cross section  $\sigma$ at 13 TeV center of mass energy, with QCD
correction of $K=1$, and eq.\ 
(\ref{eq:bestfit}) to determine $g_\L$, we find the product $\sigma B$ versus 
$m_{Z'}$ shown in fig.\ \ref{dilepton}. 

The published ATLAS limit applies to models in which equal numbers of
electrons and muons are produced.  In our model, since primarily muons
are produced, and the efficiency for detecting electrons is greater
than for muons, the limit is relaxed.  In the most interesting mass
bin for our purposes, 3-6 TeV, the relative efficiency  for electron
versus muon detection is $r=0.45/0.32=1.4$.  The bound on $\sigma B$ is then relaxed
by the factor $(1+r)/(1+r/4) = 1.8$ \cite{Moenig}, using $B(ee) = B(\mu\mu)/4.$
This leads to the limit $m_{Z'}>4.3\,$TeV, which when combined with
(\ref{eq:bestfit}) implies a gauge coupling $g_\L \gtrsim 0.7$.  Thus
another prediction of this model is that the $Z'$ should appear soon
in searches for resonant dimuons, if the gauge coupling is not 
much greater than $\sim 1$.

Recently a complementary recasting of dilepton constraints was done
in ref.\ \cite{Greljo:2017vvb}, pointing out that they could also
limit the size of effective 4-fermion operators induced by integrating
out a heavy $Z'$, even if its mass is beyond the reach of resonant
production at LHC.  Coefficients of the operators 
$(\bar Q_1\gamma^\mu Q_1)(\bar L_{e,\mu}\gamma_\mu L_{e,\mu})$
involving first generation left-handed quarks and 
contributing to $pp \to e^+e^-$ and $pp\to \mu^+\mu^-$ are bounded
using the resonant dilepton searches.  The {dimensionless coefficients} 
are identified in our model and constrained as
\bea
	C^{(1)}_{Q_1 L_e} &=& -{ g_\L^2 v^2\over {6}\, m_{Z'}^2} = 
	(-1.8\pm0.4)\times 10^{-4}\nonumber\\
	& \notin& [0.0,\,1.75]\times 10^{-3}\nonumber\\
	C^{(1)}_{Q_1 L_\mu} &=& {2 g_\L^2 v^2\over {3}\, m_{Z'}^2} = 
	(7.2\pm1.9)\times 10^{-4}\nonumber\\	
	& \in& [-5.73,\,14.2]\times 10^{-4}
	\label{eq:boundstails}
\eea
where the $2\,\sigma$ allowed ranges are given.  There is some tension
at the level of $\sim 2\,\sigma$
in the $pp\to e^+e^-$ channel, where the range is asymmetric
because of a deficit of background events in some invariant mass bins. 
This analysis reinforces the conclusion that dilepton searches could
soon reveal evidence for our model, or exclude it.

Interestingly, an independent constraint on  the
$C^{(1)}_{Q_1 L_e}$ Wilson coefficient arises from parity-violating
observables in atomic and electron-proton scattering experiments:
$C^{(1)}_{Q_1 L_e} =(1.6\pm1.1)\times
10^{-3}$~\cite{Falkowski:2017pss}, which is consistent with the bound
in eq.~(\ref{eq:boundstails}). There is no analogous constraint on
$C^{(1)}_{Q_1 L_\mu}$,  but the muonic coupling of the $Z^\prime$ can
be tested using neutrino trident production
\cite{Altmannshofer:2014pba}, which in our case leads to the lower
limit $M_{Z^\prime}/g_L\gtrsim 700$ GeV.

%---------------------------------------------------------------------------------------------------
%
%
%---------------------------------------------------------------------------------------------------

\bigskip
\subsection{$Z$ and $W$ couplings to fermions}
\label{sec:WZ}

As we dicuss in detail below, in order to have a $Z^\prime$ with a 
mass of $\sim5$ TeV, at least some of 
the exotic fermions $U$, $D$, $E$ and $N$ must be in the multi-TeV range. 
Although such masses are still out of reach for direct searches at 
the LHC, they can affect low-energy observables like the couplings 
of weak bosons to the SM fermions. By integrating out all the heavy states in eq.~(\ref{eq:model}), at the electroweak scale they produce the effective operators
\begin{align}
\label{eq:LageffHF}
\mathcal L_{\rm eff}\supset&-\frac{1}{4}\left(\frac{\lambda_d^2}{{ M}_d^2}-\frac{\lambda_u^2}{{ M}_u^2}\right)(H^\dagger i\,\raisebox{2mm}{$\nada^\leftrightarrow$} \hspace{-4mm} D_\mu\,H)(\bar Q_\L\gamma^\mu Q_\L)\nonumber\\ 
& -\frac{1}{4}\left(\frac{\lambda_d^2}{{ M}_d^2}+\frac{\lambda_u^2}{{ M}_u^2}\right)(H^\dagger i\,\raisebox{2mm}{$\nada^\leftrightarrow$} \hspace{-4mm} D_\mu^I\,H)(\bar Q_\L\gamma^\mu\tau^I Q_\L)\nonumber\\
&-\frac{1}{4}\left(\frac{\lambda_\ell^2}{{ M}_\ell^2}-\frac{\lambda_\nu^2}{{ M}_\nu^2}\right)(H^\dagger i\,\raisebox{2mm}{$\nada^\leftrightarrow$} \hspace{-4mm} D_\mu\,H)(\bar L_\L\gamma^\mu L_\L)\nonumber\\
&-\frac{1}{4}\left(\frac{\lambda_\ell^2}{{ M}_\ell^2}+\frac{\lambda_\nu^2}{{ M}_\nu^2}\right)(H^\dagger i\,\raisebox{2mm}{$\nada^\leftrightarrow$} \hspace{-4mm} D_\mu^I\,H)(\bar L_\L\gamma^\mu\tau^I L_\L),
\end{align}
where $\raisebox{2mm}{$\nada^\leftrightarrow$} \hspace{-3.5mm}
D_\mu=D_\mu-\raisebox{2mm}{$\nada^\leftarrow$} \hspace{-3mm} D_\mu$ and
$\raisebox{2mm}{$\nada^\leftrightarrow$} \hspace{-3.5mm} D_\mu^I=\tau^ID_\mu-
\raisebox{2mm}{$\nada^\leftarrow$} \hspace{-3mm} D_\mu \tau^I$,
both of which act trivially in generation space.~\footnote{We have
omitted the contributions to operators of the type $(H^\dagger H)(\bar Q_L H d_R)$, \ldots which are also generated by integrating out the heavy
fermions but that are only weakly constrained by Higgs-couplings 
measurements.}\ \ These can readily be converted into modifications of the $Z$ and $W$ 
couplings of the SM left-handed fields,
\begin{align}
\label{eq:coupsweak}
\delta g_{\Z,\L}^{f}=-\frac{v^2\,\lambda_{f}^2}{{ M}_{f}^2},~~~\delta
g_\W^{\rm q,\ell}=-\frac{v^2}{2}\left(\frac{\lambda_{d,\ell}^2}{{ M}_{d,\ell}^2}+\frac{\lambda_{u,\nu}^2}{{ M}_{u,\nu}^2} \right).
\end{align}

The $Z$ couplings to the fermions have been measured very precisely
at  LEP. The strongest constraint is on the coupling to the leptons,
$\delta g_{Z,L}^\ell=(-0.0952\pm0.215)\times10^{-3}$ which leads to
the bound ${ M}_\ell/\lambda_\ell\geq 7.7$\,TeV at 95\%
C.L.~\cite{Falkowski:2017pss}. The invisible width of the $Z$ leads to the bound $\delta g_{Z,L}^\ell=(-1.32\pm0.72)\times10^{-3}$
or $M_\nu/\lambda_\nu>3.3$\,TeV at 95\% C.L. 
Similarly, for the couplings to the up- and down-type quarks we get ${
M}_u/\lambda_u\geq 2.4$\,TeV and ${ M}_d/\lambda_d\geq 6.1$ TeV at 95\%
C.L. 

Generally these {bounds} can be satisfied even if $M_f$ is not very large
by taking $\lambda_f$ sufficiently small. The exact formula for the 
{heavy fermion masses, eq.~(\ref{eq:exactmassexp}),} implies 
that $\lambda_f > m_f/v$ (since
$\lambda'\Phi/\sqrt{M^2+(\lambda'\Phi)^2} <1$), where $m_f$ is the 
mass of the heaviest SM fermion of type $f$.  Thus for down-type
quarks, the $Z$ decay bound can be satisfied even if $M_d = 21\,$GeV.
However for the up-type quarks we have $\lambda_u\gtrsim 1$, hence
$M_u\geq 2.4\,$TeV. 
These constraints can be expressed in terms of the couplings
$\lambda_f$, $\lambda''_f$ by using eqs.\ (\ref{gbmass}) and 
(\ref{eq:bestfit}) 
(see the discussion around eqs.\ (\ref{mZpeq}) and (\ref{Mfeq})),
resulting in upper limits 
shown in table \ref{Ztab}.

\begin{table}[t]
\begin{tabular}{|c|c|c|c|c|}
\hline
$f = $ & $u$ & $d$ & $\ell$ & $\nu$ \\
\hline
${\lambda_f/\lambda''_f}  < $ & {$2.2$} & {$0.87$} & {$0.69$} & {$1.6$} \\
\hline
\end{tabular}
\caption{Upper limit on $\lambda_f/\lambda''_f$ {at 95\% C.L.} from LEP constraints
on $Z\to f\bar f$ decays.}
\label{Ztab}
\end{table}

In the case of the charged currents, the strongest bound stems from the first-row unitarity test of the CKM matrix~\cite{Gonzalez-Alonso:2016etj},
\begin{align}
\label{eq:CKMunit}
\Delta_{\rm CKM}=&|\tilde V_{ud}|^2+|\tilde V_{us}|^2+|\tilde V_{ub}|^2-1\nonumber\\
\simeq&v^2\left(\frac{\lambda_{\ell}^2}{{ M}_{\ell}^2}+\frac{\lambda_{\nu}^2}{{ M}_{\nu}^2}-\frac{\lambda_{d}^2}{{ M}_{d}^2}-\frac{\lambda_{u}^2}{{ M}_{u}^2}\right),
\end{align}
where in the second line we have used the corrections in 
eq.~(\ref{eq:LageffHF}). The experimental bound is
$\Delta_{\rm CKM}=(-4.2\pm5.2)\times 10^{-4}$, while the contributions
to (\ref{eq:CKMunit}) from the charged leptons can be as large as 
$(v/7.8\,{\rm TeV})^2\cong 5\times 10^{-4}$, from the $Z$-decay
constraint.  If the other contributions are no larger
(even though the $Z$ decay bounds would allow
them to be so), the constraint is satisfied without any need for
tuning
of parameters.  This is the case if {$\lambda_f/\lambda''_f \lesssim
\sqrt{5\times 10^{-4}}\,\langle{\cal M}\rangle/\,v \cong 0.68$},
which is consistent with the $Z$ decay limits in table \ref{Ztab}.

\section{UV implications}
\label{sec:UV}

The discussion so far has been focused on explaining the
$R_{K^{(*)}}$ anomalies while satisfying 
other flavor-changing constraints on the low-energy
limit of the theory.  Here we return to the higher-energy regime
to explore how this relates to the masses of the heavier gauge bosons,
and the mechanism of fermion mass generation.
\bigskip
\subsection{Hierarchy of scales}
\label{sec:hier}

We require the octet scalars $\Phi_{8,i}$ 
to get VEVs
proportional to the generators $T^{1,2}$ in order to give large 
masses
to all the $Z's$ that couple to generators other than $T^8$.
Supposing that $\langle\Phi_8\rangle = \alpha T^1$,
$\langle\Phi_{8,2}\rangle = \beta T^2$ and 
no other VEVs are present, the gauge boson mass matrix is
\be
	M^2_{\rm gb} = g_\L^2 \,{\rm diag}
	\left(\beta^2,\,
\alpha^2,\,\gamma^2,\,\sfrac14\gamma^2,\,
\sfrac14\gamma^2,\,\sfrac14\gamma^2,\,\sfrac14\gamma^2,\,0\right)
\label{mdiag}
\ee
where $\gamma^2=\alpha^2+\beta^2$. Separate contributions from two
octet fields are required to avoid a second vanishing eigenvalue,
that would lead to large FCNC's amongst light quarks,
The first two of the states in (\ref{mdiag}) couple to $T^{1,2}$, which mediate
$s\to d$ transitions.  Constraints from $K$-$\bar K$ mixing require
that $g_\L\alpha,\ g_\L\beta \gtrsim 10^4$\,TeV, since their
exchange generally produces $(\bar d_\L\gamma^\mu s_\L)^2$ with
a coefficient whose imaginary part is not suppressed. 
(Rotation $T^{1,2}\to V_{\rm CKM}^\dagger T_{1,2} V_{\rm 
CKM}$ to the quark
mass basis does not affect this conclusion).

Since $T^8$ commutes with $\langle\Phi_{8,i}\rangle$, only the  second
term in (\ref{fmass}) contributes to the $Z'$ mass. 
Recalling the
simplifying assumption that the ${\cal M}$ VEV is proportional to
the unit matrix, we have 
\be
\label{mZpeq}
   m^2_{Z'} = {g_\L^2} \langle{\cal M}\rangle^2. 
\ee 
Then using (\ref{eq:bestfit}) the heavy fermion masses are given by
\be 
\label{Mfeq}
   M_f  = \lambda''_f\, \langle{\cal M}\rangle  = \lambda''_f \times 5.3\,{\rm TeV} 
\ee 
Assuming the couplings
$\lambda_f''\lesssim 1$, this implies that all the heavy fermions
are within the reach of the LHC.
Current limits on vectorlike quark masses are still close to 1 TeV 
\cite{Sirunyan:2017usq,Aaboud:2017qpr}.

\subsection{$R_D$, $R_{D^*}$}
It is interesting to ask whether the present framework could also
accommodate the anomalies observed in the decays $B\to
D^{(*)}\tau\nu$.  It would require the presence of a heavy $W'$ boson
in addition to the $Z'$.  In principle this could be accomplished
by extending the gauge symmetry to SU(6)$_\L\times$SU(3)$_\R$,
where SU(6)$_\L$ contains the SM gauge group SU(2)$_\L$.  The
additional $W'$ gauge bosons then arise from the breaking of
SU(6)$_\L\to$ SU(2)$_\L\times$SU(3)$_\L$.  
CKM-like mixing would produce the generation-changing interaction
\be
	{g_\L^2 V_{cb}\over 4 m_{W'}^2} 
	(\bar c_\L\gamma^\mu b_\L)(\bar\tau_\L
	\gamma_\mu\nu_\L).
\label{RD}
\ee 
Although such an operator can provide a good fit to $R_{D^{(*)}}$,
there are two problems in the present framework.  First, eq.\
(\ref{ltrans}) also predicts the operator 
$(\bar c_\L\gamma^\mu b_\L)(\bar\mu_\L	\gamma_\mu\nu_\L)$ with
coefficient $-2$ times that in (\ref{RD}), which does not fit the
observations~\cite{Glattauer:2015teq,Abdesselam:2017kjf}.  Secondly, as 
shown in ref.\ \cite{Freytsis:2015qca}, 
the required mass for $W'$ to fit $R_{D^(*)}$ is too small to satisfy
LHC constraints, given that the $W'$ couples to light quarks through
the generator (\ref{qgen}). 

\subsection{Asymptotic freedom}
With the particle content listed in table \ref{charges}, 
the $g_\L$ and $g_\R$ couplings remain asymptotically free.
The contributions to the $\beta$ functions are 
\bea
\label{betaL}
	{16\pi^2\over g_\L^3}\beta(g_\L) &=&
	-11 + {16\over 3} + \frac12 + 1 = -{25\over 6}\\
\label{betaR}
	{16\pi^2\over g_\R^3}\beta(g_\R) &=&
	-11 + {16\over 3} + \frac12 + 2 + {5\over 6} = -{7\over 3}
\eea
from the gauge bosons, fermions, bifundamental scalars, 
and octets (plus sextet in (\ref{betaR})), respectively.  We have assumed that the $\Phi_f^A$
component fields are real, so that the matrices $\Phi_f$ and hence
the SM Yukawa matrices are 
Hermitian, which is phenomenologically allowed \cite{Evans:2011wj}.

The beta function (\ref{betaL}) is only valid above the scale
$\langle\Phi_{8,i}\rangle\sim 10^4\,$TeV at which SU(3)$_\L$ is 
restored.  Between this scale and $m_{Z'}$, we should consider the
evolution of $g_\L$ as the gauge coupling of the U(1) associated with
$Z'$.  Its beta function is given by
\be
	\beta(g_\L) = {g_\L^3\over 12\pi^2}\times \left(4 + \frac38
\right)
\ee
where the respective contributions from the fermions and bosons
are shown.  Using the initial condition $g_\L=0.7$ at $\mu =
m_{Z'}=4.2\,$TeV, which would saturate the current bound from ATLAS
dilepton searches, this would lead to a Landau pole at scale
$\mu \sim 10^{12}\,$TeV.  However asymptotic freedom takes over well
before, at $10^4\,$TeV, so the theory has good UV behavior.

\subsection{Neutrino masses}
A further consequence of the structure of the currents is that we
are forced to take the transformation $V_l^\L$ that
diagonalizes the lepton mass matrix to be close
to the permutation (\ref{Vleq}).  This means that the lepton masses
have to be in an unusual order in the original basis, 
diag($m_\mu,m_\tau,m_e$).  As mentioned above, this fixes the 
left-handed neutrino rotation in terms of the PMNS matrix $U$ to be
\be
V_\nu^\L =  U V_l^\L \cong\left(\begin{array}{ccc}
-0.15& 0.82& 0.54\\
 0.62& -0.35& 0.70\\
 0.77& 0.44& -0.45
	\end{array}\right)
\ee
From this one can infer the form of the seesaw neutrino mass matrix
in the original basis, before diagonalization:
\bea
	\tilde m_\nu &\cong& m_{\nu_3}\left(\begin{array}{ccc}
	0.65 & 0.30 & -0.28\\
	0.30 & 0.21 & -0.24\\
	-0.28 & -0.24 & 0.28
	\end{array}\right),\quad \hbox{normal}\nn\\
&\cong& m_{\nu_3}\left(\begin{array}{ccc}
	0.98 & 0.12 & 0.09\\
	0.12 & 0.32 & -0.44\\
	0.09 & -0.44 & 0.69
	\end{array}\right),\quad \hbox{inverted}\nn
\eea
depending upon whether the mass hierachy is normal or inverted.
We assumed that $m_{\nu_1}\ll m_{\nu_{2,3}}$.

\section{Discussion} 
\label{sec:conc}

Our model has similarities to that of ref.\ \cite{Alonso:2017bff}, in
which the gauged flavor symmetry is SU(3)$_H\times$U(1)$_{B-L}$ acting
vectorially on the SM fermions.  The horizontal symmetry is the same
as we have considered except for the fact that it is not chiral and
it includes an extra U(1) factor.  This leads to a number of 
important phenomenological differences between the models.  First,
right-handed currents are present in ref.\ \cite{Alonso:2017bff}
(though they are taken to be flavor-diagonal), while they are presumed
to be negligible in ours.  Second, the flavor generators in
\cite{Alonso:2017bff} are not traceless like in eqs.\  (\ref{qgen})
and (\ref{ltrans}).  Third, since $B-L$ is opposite for quarks and
leptons, the currents for quarks and leptons are different linear
combinations of $T^8$ and $1$ in \cite{Alonso:2017bff}, whereas they
are the same in our model. The presence of the U(1)$_{B-L}$ factor in
\cite{Alonso:2017bff} leads to a Landau pole at scales $\sim
10^{10}\,$GeV, which is not present in our model.  Fourth, our model
requires no charged lepton flavor violation, whereas it is essential for
generating the coupling to muons in \cite{Alonso:2017bff}.  Moreover,
we have {explored} 
the connection  between flavor
symmetry breaking and the Yukawa matrices of the SM in our framework.

One consequence of the tracelessness of our generators has already
been noted: new contributions to the  decays $B\to K\nu\bar\nu$ or 
$B\to \pi\nu\bar\nu$ are negligible, because the interference with the
SM contribution vanishes.  Another is that sizable couplings of $Z'$
to all three generations cannot be avoided. In ref.\
\cite{Alonso:2017bff}, VEVs for the fundamentals that break
SU(3)$_H\times$U(1)$_{B-L}\to$U(1)$_h$ are chosen such that the
leptonic generator couples only to the third generator, before mass
mixing.  By assuming the mixing is small, the branching ratio of 
$Z'\to\mu\mu$ (and even more so $Z'\to ee$) can be suppressed, making
it easier to satisfy ATLAS constraints on resonant dilepton
production.  Our model does not have this option, leading to mild
tension in this observable. The traceless generators also imply that
no gauge kinetic mixing  will arise between the $Z'$ and the SM U(1)
hypercharge at one loop. Hence potentially strong constraints from
diboson production  \cite{Osland:2017ema} are evaded in our model.

Although phenomenologically complete, our analysis does not address
how difficult it might be to construct a potential for all the scalar
fields that leads to the desired pattern of VEVs, or perhaps a similar
one that is nevertheless viable.  This is probably challenging, and might
best be postponed pending further experimental evidence in favor of
the model.  There are a number of new physics signals that should be
close to being observable, in addition to direct production of the
$Z'$ at LHC.  These include a positive contribution to the $\epsilon_K$
parameter for $K$-$\bar K$ mixing, a negative contribution to the
first-row CKM unitarity test (\ref{eq:CKMunit}), an enhancement of
the decay width for $Z\to\ell\ell$, and vectorlike quarks and leptons
at the few-TeV scale.

\bigskip {\bf Acknowledgments.}  We thank {Rodrigo Alonso, {Mattia Dalla Brida, Luca Di Luzio}, Martin
Gonzalez-Alonso}, Ben Grinstein, David London, David Marzocca, Klaus
M\"onig, Guy Moore, {Rui-xiang Shi} and Alfredo Urbano for valuable advice and discussions.

\appendix 

\section{$L$-violating decays}
\label{app:LV}

Nothing forbids the entries $\epsilon_i$ in the leptonic currents
(\ref{ltrans}), which are constrained by lepton-flavor violating
decays such as $\mu\to 3e$, $\tau\to 3l$ at the level of $10^{-12}$
and $10^{-8}$ in the respective branching ratios.  By comparing the
NP and SM Wilson coefficients for the exotic decays versus the
allowed ones, we find that 
\bea
|\epsilon_{1}| &\lesssim& 
	10^{-6}{2\sqrt{2}\,G_F\over 2\,g_\L^2/(12 m_{Z'}^2)} = 0.0067
	\nn\\
|\epsilon_{2,3}| &\lesssim& 
	10^{-4}{2\sqrt{2}\,G_F\over 2\,g_\L^2/(12 m_{Z'}^2)} = 0.67
\label{epslim}
\eea
using eq.\ (\ref{eq:bestfit}).

At one loop, these couplings also give rise to $\mu\to e\gamma$
and $\tau\to l\gamma$, through the transition magnetic moment operator
$\mu_{ij} (\bar l_\L^j [\slashed{q},\slashed{A}]l_\R^i) $, where
$q^\mu$ is the photon momentum.  We find that
\be
	\mu_{ij} = {e\,g_\L^2\over 384\,\pi^2\, m_{Z'}^2}
	\left\{\begin{array}{ll} -\epsilon_1\,m_\mu 
	\ln{m_{Z'}^2\over m_\mu^2},& \mu\to e\gamma\\
	 2\,\epsilon_2\,m_\tau 
	\ln{m_{Z'}^{2^{\phantom{|}}}\over m_\tau^2},& \tau\to e\gamma\\
	-\epsilon_3\, m_\tau 
	\ln{m_{Z'}^{2^{\phantom{|}}}\over m_\tau^2},& \tau\to\mu\gamma
	\end{array}\right.
\label{transm}
\ee
Using the decay width $\delta\Gamma_{ji} = |\mu_{ij}|^2 
(m_i^2-m_j^2)^2/
(8\pi m_i)$, and the PDG limits \cite{PDG} on the radiative decays,
 we find weaker limits than in (\ref{epslim}):
\be
	|\epsilon_1| < 0.011,\quad |\epsilon_2| < 4.2,\quad 
	|\epsilon_3| < 5.1
\ee
where we took $m_{Z'}= 6\,$TeV to evaluate the logarithms.

The muon anomalous magnetic moment is related to the $\mu\to e\gamma$
transition moment in (\ref{transm}) by taking $\epsilon_1\to 8$.  This gives a 
contribution to $(g-2)_\mu/2= 4\times 10^{-11}$, smaller than the observed
discrepancy by a factor of 75.

Lepton flavor violating decays of vector mesons, for example $J/\psi \to\mu
e$, have branching ratios of order $|\epsilon_i g_\L^2 m^2_{J/\psi}/
32 e^2 m_{Z'}^2|^2 \lesssim 10^{-15}|\epsilon_i|^2$,
far below current limits $\sim 10^{-7}$.  Pseudoscalar mesons have
chirality-suppressed decays to purely leptonic final states.
The perturbation to the branching ratio of $B_d\to\mu\mu$ is
predicted to be $\delta B/B \cong \sqrt{2 m_B/\pi\Gamma}\,(V_{td}\,g_\L^2f_B m_\mu)/(12
m_{Z'}^2)\cong 0.08$, which is smaller than the experimental 
error of $0.4$.  For the $L$-violating decays such as $B_{s,d}\to\mu e$,
since there is no interference with the SM the predicted signal is
much smaller and gives no useful limits on $\epsilon_i$. 

\section{Vectorial flavor symmetry}
\label{app:vector}  

One could imagine constructing a similar model to the one we have
proposed, but using a vectorial SU(3)$_\H$ flavor symmetry instead of
SU(3)$_\L\times$SU(3)$_\R$.  The same interactions as in eq.\ 
(\ref{eq:model}) could be written, but the fields ${\cal M}_f$ would
have to be in the $8$ representation rather than bifundamental, and a
discrete symmetry would be required to forbid large flavor-universal
contributions to the Yukawa matrices involving only SM fields.  The
flavor-conserving quark and lepton currents would be vectorial, while
the FCNCs of the quarks would be left-handed as in the model of ref.\
\cite{Alonso:2017bff}.  A good fit to $R_{K^{(*)}}$ can still be
obtained with vectorial leptonic currents; some authors would argue
that this is even preferred~\cite{Capdevila:2017bsm}.  

There are several major drawbacks however.  First, the sextet field
cannot get a large VEV to produce heavy right-handed neutrino masses
while leaving a relatively light $Z'$, making the origin of neutrino
masses problematic.  (The extra U(1)$_{B-L}$ factor allowed  ref.\
\cite{Alonso:2017bff} to overcome this problem.) Second, the tension
with  dilepton searches is multiplied by having vectorial couplings to
the $Z'$.  For resonance searches, the production cross section
increases by a factor of 2, while for the nonresonant constraints the
number of operators simultaneously contributing to the signal with
equal strength is quadrupled, creating a significant tension in all
channels but especially electrons.   Finally, asymptotic freedom of
the gauge coupling is badly spoiled by the large matter content,
including 10 octet scalars and a heavy copy of the SM fermions,
leading to a Landau pole at a relatively low scale.

\bibliographystyle{apsrev}
%\bibliography{ref_abelian_dm,scalarDM,seesaw,ref4thGen}

\end{document}